\documentclass[twocolumn,showpacs,preprintnumbers,amsmath,amssymb]{revtex4}

\usepackage{graphicx}
\usepackage{dcolumn}
\usepackage{bm}
\usepackage{amssymb}

\def\address{\@ifstar{\address@star}%
  {\@ifnextchar[{\address@optarg}{\address@noptarg}}}

\begin{document}

\author{S.N.~Gninenko$^{1,2}$}
\author{N.V.~Krasnikov$^{1}$}
\author{A.~Rubbia$^{2}$}

\affiliation{$^{1}$Institute for Nuclear Research of the Russian Academy of Sciences, Moscow 117312}
\affiliation{$^{2}$Institut f\"ur Teilchenphysik, ETHZ, CH-8093 Z\"urich, Switzerland}


\title{Search for  millicharged particles in reactor neutrino experiments: \\
a probe of the PVLAS anomaly}

\date{\today}

\begin{abstract}

It has been recently suggested that the vacuum magnetic dichroism observed by 
the PVLAS experiment   could be explained by 
pair production of a new light, $m_\varepsilon\simeq 0.1$ eV,
  millicharged, $q_\varepsilon \simeq 3\cdot 10^{-6} e$, fermions 
($\varepsilon$).
In addition, it has been  pointed out that  millicharged particles 
with $q_\varepsilon \gtrsim  10^{-9} e$ appear naturally  in models based on 
the string theory.

We show that low energy
reactor neutrino experiments  
provide a sensitive probe of   
millicharged particles. Considering, as an example,  recent results of the
TEXONO experiment searching for neutrino magnetic moment,  
a new upper bound  $q_\varepsilon \lesssim 10^{-5} e$ 
 for the mass region $m_\varepsilon < 1$ keV is derived. 
These results enhance motivations for  a more sensitive search for 
such particles in near future experiments.
 Furthemore, a  direct experimental limit on 
the electric charge of the electron antineutrino $q_{\bar{\nu}_e} < 3.7 \cdot 10^{-12} e$ is obtained.
\end{abstract}
\pacs{14.80.-j, 12.20.Fv, 13.20.Cz}
\maketitle

Several realistic extensions of the standard model of elementary particle physics suggest 
the existence of particles with small, unquantized electric charge. 
They appear naturally and  generically from kinetic-mixing in theories 
containing at least two U(1) gauge factors, e.g. in
paraphoton models \cite{okun}. By adding a second,
hidden photon (the ``shadow'' photon) one could construct
grand unified models which contain particles with an electric
charge very small compared to the electron charge \cite{holdom}.
These considerations have stimulated new theoretical works 
and experimental tests,
see e.g.  \cite{moh1,dobr,slac,davidson,dub} and references therein.\\

Recently, a laboratory experiment PVLAS has reported on observation of an
anomalous rotation of the polarization plane of a 
photon beam incident on a region of a static transverse  magnetic field in 
vacuum \cite{pvlas}. 
The PVLAS signal appears to be in contradiction with the negative 
results of the CAST experiment on search for solar axions \cite{cast} and
astrophysical constrains \cite{raffelt} (however, 
one may argue that  astrophysical
constraints  are also model dependent and have
considerable theoretical and experimental uncertainties).

Although many models explaining the PVLAS-CAST discrepancy    
appeared recently \cite{papers}, not all of them 
allow  for experimental tests.
One of the experimentally interesting explanation suggests that the PVLAS   
anomalously large effect 
 may originate from pair production of 
light, $m_\varepsilon\simeq 0.1$ eV, millicharged, $q_\varepsilon \simeq 3\cdot 10^{-6} e$, fermions ($\varepsilon$), 
whose  existence  is not excluded by the present experimental 
data \cite{gies}. 
It is also pointed out that 
the  multiple U(1) factors, 
the size of kinetic-mixing, and suitable matter representations required 
to explain the 
PVLAS data occur very naturally in the context of realistic extensions of the 
Standard Model  based on string theory \cite{ring}. The string scale $\simeq 10^{11}$ GeV 
predicts the existence of particles with $q_\varepsilon\gtrsim 10^{-9}e$, 
thus making the  region $q_\varepsilon \simeq 10^{-9} - 10^{-5} e$  of great interest
for searching for millicharged particles in future
 experiments.\

In this note, we  show that
reactor neutrino experiments
searching for non-standard neutrino interactions  provide
 a sensitive probe to light millicharged particles.
We discuss possibilities for more sensitive searches 
for such particles in near future experiments.

It is well known that since the electromagnetic cross section 
is orders of magnitude larger than the weak cross section, the  existence
of    non-zero magnetic moment of the electron antineutrino would
increase  the total rate of events in  
$\bar{\nu} - e^-$- scattering experiments \cite{vogel}. 
The integral cross section for neutrino magnetic scattering $\sigma_{\mu}$ 
depends very weakly  on the neutrino energy. It
rises only logarithmically with the neutrino energy, while the total neutrino
cross section rises linearly with neutrino energy $E_{\nu}$.
 Thus, it is advantageous to
search for neutrino magnetic moment using low neutrino energy. 
These considerations point to nuclear reactors as probably the most
convenient source of neutrinos. Reactors with power of 2 GW 
emit  about $4\times 10^{20}~\bar{\nu}_e$ per second \cite{bemp}. The emission 
energy spectrum is quite well understood. It has a 
broad distribution over energies up to 
10 MeV, with a maximum intensity at 0.5-1.5 MeV \cite{vogel}.  

Consider, as an example, recent results of the TEXONO collaboration
 searching for
electron anti-neutrino magnetic moment \cite{texono}. 
The experiment has been set up at the Kuo-Sheng Nuclear Power Station at a distance 
of 28 m from the 2.9 GW reactor core. 
A detector threshold of 5 keV and a background level of 1 counts day$^{-1}$keV$^{-1}$kg$^{-1}$ 
at 12-60 keV was achieved with a high purity germanium detector of mass 
1.06 kg surrounded by anti-Compton detectors with NaI(Tl) and CsI(Tl) crystal 
scintillators. 

The excess $\Delta N_{\nu-e}$ of events in the Ge detector
due to the $\bar{\nu} -e $ magnetic scattering 
is given by
\begin{equation}
\Delta N_{\nu-e} = k \int^\infty_{E_\nu^{min}} \int^{E_e^{max}}_{E_e^{min}} \frac{dN_\nu}{dE_\nu} \frac{d\sigma_\mu}{dE_e}  dE_e dE_\nu 
\label{nueexcess}
\end{equation}
where $E_e$ is the electron recoil energy, $dN_\nu/dE_\nu$ is the reactor 
neutrino flux, $d\sigma_\mu/dE_e$ is the $\bar{\nu} - e$ scattering differential 
 cross section  via the $\nu$ magnetic moment, and 
 $k$ is a factor corresponding to the convolution of the Ge detector acceptance  and mass.

The  cross section $d\sigma_\mu/dE_e$ is given by, see e.g. \cite {vogel}:
\begin{equation}
\frac{d \sigma_{\mu}}{dE_e} = \frac{\pi \alpha^{2}}{m_{e}^{2}}
\frac{\mu_{\nu}^{2}}{\mu_{B}^{2}}\Bigl(\frac{1}{E_e} - \frac{1}{E_{\nu}}\Bigr)
\end{equation}
where 
$\mu_{\nu}$ is the neutrino magnetic moment and $\mu_B = e/2 m_e$ is Bohr magneton.

One can see that for the  
electron energy interval from  $E_e^{min}=$ 12 keV  to
$E_e^{max}=$ 60 keV,  analyzed by TEXONO \cite{texono},
the cross section of $\bar{\nu_e} - e$ scattering due to non-zero magnetic moment 
 is practically independent of energy for $E_\nu \gtrsim 0.5$ MeV
and is equal:
\begin{equation}
\sigma_{\mu} = 4.3 \times 10^{-45} \Bigl(\frac{\mu_{\nu}}{10^{-10}\mu_B}\Bigr)^2~cm^2
\end{equation}
Thus, the total number $\Delta N_{\nu-e}$  of events collected during period 
of time $\Delta t$
 could be estimated as 
\begin{equation}
\Delta N_{\nu-e} =  f_{\nu}  \sigma_{\mu}  N_e   \Delta t 
\label{nue}
\end{equation} 
where $f_{\nu} = 5.8\times10^{12}$ cm$^{-2}$ s$^{-1}$  is the integrated 
reactor neutrino flux ( the detection efficiency of the Ge detector is 
$\simeq 100\%$ for energy deposition above 10 keV) \cite{texono} and  
$N_e$ is the total number of electrons in the detector.

Using the above values we  found
that the TEXONO experiment should detect an excess of 
$\Delta N_{\nu-e}  \simeq 5.8~ events /day$ if 
$\mu_{\nu_{e}}=2\times 10^{-10}\mu_B$. 
This value  is in an excellent  agreement with  the number of events
quoted in ref. \cite{texono}.

Having this in mind, consider the case of millicharged particles.
The nuclear reactors could also be a 
source of such light particles. Indeed, the 2.9 GW power reactor core emits 
large number of $\gamma$-quanta, more than $5\times 10^{20}$ 
per second,  and these gammas 
could convert within the reactor into  
pairs of light $\varepsilon$-particles.    
Even a small fraction of these particles
 could lead to an observable excess of 
electrons from $\varepsilon-e^-$ scattering in the 
detector whose experimental  signature is identical to that of $\bar{\nu} - e^-$ magnetic 
scattering events.

Consider now the number of $\varepsilon-e$ events in the TEXONO Ge detector. 
 The excess $\Delta N_{\varepsilon-e}$ of events 
due to $\varepsilon-e$ scattering in the detector is given by:
\begin{equation}
\Delta N_{\varepsilon-e} = k \int^\infty_{E_\varepsilon^{min}} \int^{E_e^{max}}_{E_e^{min}} \frac{dN_\varepsilon}{dE_\varepsilon} \frac{d\sigma_\varepsilon}{dE_e}  dE_e dE_\varepsilon 
\label{qeexcess}
\end{equation}
where  $dN_\varepsilon/dE_\varepsilon$ is the  assumed $\varepsilon$-particles flux, 
and $d\sigma_\varepsilon/dE_e$ is the $\varepsilon-e$ scattering differential 
 cross section.

 For $E_\varepsilon  \gg E_e^{min} \gg m_\varepsilon$, i.e. $m_\varepsilon < 1$ keV,
 the  cross section $d\sigma_\varepsilon/dE_e$  is independent of $E_\varepsilon$ and 
 is given by 
\begin{equation}
\frac{d\sigma_{\varepsilon}}{dE_e} = 2\pi r_0^2 \Bigl(\frac{q_\varepsilon}{e}\Bigr)^2  \frac{m_e}{E_e^2} 
\end{equation}
where $r_0$ is the classical electron radius, or  
\begin{equation}
\sigma_{\varepsilon} = 2 \times 10^{-23} \Bigl(\frac{q_\varepsilon}{e}\Bigr)^2~cm^2
\label{qecrs} 
\end{equation}
for the  recoil energy region from  $E_e^{min}=$ 12 keV  to
$E_e^{max}=$ 60 keV.

 The highest rate mechanism for the light $\varepsilon$-particle production  
 by $\simeq $1 MeV gammas is the $\varepsilon\bar{\varepsilon}$-pair production in the  Compton scattering 
\begin{equation}
e^- + \gamma \rightarrow e^- + \gamma^* \to e^-  + \varepsilon\bar{\varepsilon}
\label{compton}
\end{equation} 
The cross section ($\sigma_{\varepsilon\bar{\varepsilon}}$) of this  process is proportional to 
$q_\varepsilon^2$ and is characterized by $q_\varepsilon, m_\varepsilon$ and spin of the 
$\varepsilon$-particles. 
It dominates 
over Bethe-Heitler pair production which is proportional to $q_\varepsilon^4$.   

For the given photon energy interval ($E_\gamma$; $E_\gamma + d E_\gamma$) 
the total number of $\varepsilon$-particles per second produced in the reactor 
core can be calculated as:
\begin{equation}
d N_\varepsilon = 2 \frac{dN_\gamma}{dE_\gamma}  \sigma_{\varepsilon\bar{\varepsilon}} \rho N_A \frac{Z}{A}  L  d E_\gamma 
\label{qnum}
\end{equation}
where $dN_\gamma/dE_\gamma$ is the reactor gamma spectrum and $L$ is the mean free path of photons
in the reactor core given by
\begin{equation}
L= \frac{A}{ Z \sigma_{tot}  N_A  \rho}
\label{mean}
\end{equation}
where $\sigma_{tot}, N_A, A, Z,  \rho$ are the total photon cross section,
the Avogadro's number, the atomic weight and  charge, and the density of 
the core material (we assume for the moment that this material has uniform composition). The cross section  $\sigma_{\varepsilon\bar{\varepsilon}}$ is 
found to be  
\begin{equation}
\sigma_{\varepsilon\bar{\varepsilon}} \simeq \frac{\alpha}{3\pi} \sigma_C\Bigl(\frac{q_\varepsilon}{e}\Bigr)^2 ln\frac{E_\gamma}{2m_\varepsilon} 
\label{compt}
\end{equation}
where $\sigma_C$ is the Compton scattering cross section.
It  is calculated under assumption $\sigma_{\varepsilon\bar{\varepsilon}} \ll \sigma_C$
and is valid for $\alpha/3\pi (q_\varepsilon/e)^2 ln(E_\gamma/2m_\varepsilon) \ll 1  $.

Although  the reactor gamma spectrum has all possible energies up to 10 MeV, not 
all photons  
effectively contribute to the $\varepsilon\bar{\varepsilon}$ pair production through the reaction of Eq.(\ref{compton}).
Indeed, for high $Z$-materials, $Z\simeq 90-100$, in the low energy region $E_\gamma \lesssim 1$ MeV 
the total gamma cross section is dominated by the 
atomic photoelectric effect, while at energies higher than $E_\gamma \gtrsim 5$ MeV  the 
$e^+ e^-$ pair production dominates.
Thus , the suitable gamma energy interval for the effective $\varepsilon\bar{\varepsilon}$ pairs production 
via reaction of Eq. (\ref{compton}) 
is 1 - 5 MeV.
 For gammas with these energies the total cross section is dominated by the 
Compton scattering  for any material with Z up to 100. Hence,   
we may assume $\sigma_{tot} \simeq \sigma_C$.
Combining this with Eq.(\ref{qnum}), Eq.(\ref{mean}) and Eq.(\ref{compt})
 one can see that the 
total number of $\varepsilon$'s produced inside the reactor per sec is 
\begin{equation}
N_\varepsilon \simeq \frac{2 \sigma_{\varepsilon\bar{\varepsilon}}}{\sigma_C} N_\gamma = \frac{2 \alpha}{3\pi} \Bigl(\frac{q_\varepsilon}{e}\Bigr)^2ln\frac{E_\gamma}{2m_\varepsilon} N_\gamma
\label{qtot}
\end{equation}
Thus, in this approximation  the $\varepsilon$-flux is   
independent on fuel composition,  reactor core geometry,  etc.

The total number of photons produced in reactor core can be estimated by using results 
on  $\gamma$-energy spectra measurements
\cite{reactor},  used in the experiment
searching for axion decays at a nuclear power reactor at Bugey 
\cite{bugey}. 
The reactor gamma spectrum receives contributions from several processes 
following the nuclear fission process, see e.g. \cite{bugey}. The comparison 
of numbers of neutrinos \cite{vogel} and gammas \cite{reactor}
 emitted per fission for 
various energy intervals shows that the 
total  number of   gammas 
emitted in the energy range 1 - 5 MeV
 for the nuclear fuels 
of interest, i.e. $^{235}$U, $^{239}$Pu, $^{238}$U, $^{241}$Pu, and $^{252}$Cf,
is comparable with the corresponding number of neutrinos. For example, for 
 $^{235}$U the number of neutrinos and gammas emitted in the 1-5 MeV  energy 
range  is found to be 
$N_\nu = 5.7$  and $N_\gamma = 2.1$  per fission, respectively.
The ratio of the number of neutrinos to the number of gammas are 
the same within $\pm 15\%$ for different type of fuels. 
For the energy range 0.15 - 0.5 MeV the number of gammas is  
 about 30-40\% higher than the number of neutrinos. 
The energy interval 5-10 MeV is not very important, because 
less than about 5\% of particles have those energies.
Conservatively,  we may assume that $N_\gamma  = 0.3 N'_\nu$ of the 
neutrino flux $N'_\nu$ in the energy interval 1- 5 MeV, or 
\begin{equation}
N_\gamma = 0.2 N_\nu
\label{gtot}
\end{equation}
of the total neutrino flux $N_\nu$. This is calculated by using 
 reactor neutrino spectra for different nuclear fuels  described  in 
ref. \cite{vogel}. 

Similar to Eq.(\ref{nue}) the total number $\Delta N_{\varepsilon-e}$  of $\varepsilon-e$
events collected during period 
of time $\Delta t$ is 
\begin{equation}
\Delta N_{\varepsilon-e} =  f_\varepsilon   \sigma_\varepsilon N_e  \Delta t 
\label{qe}
\end{equation} 
where $f_\varepsilon$  is the flux of $\varepsilon$'s with energy $E_\varepsilon \gg E_e^{min}$.
The energy of $\varepsilon$'s produced by 1- 5 MeV gammas  is found to be widely distributed 
over the  range $0<E_\varepsilon < 5$ MeV with a few \%'s of events lying in the low energy 
region $E_\varepsilon \lesssim 50$ keV.   

Combining Eq.(\ref{qtot}) and Eq.(\ref{gtot}), the flux of $\varepsilon$'s per sec is 
\begin{equation}
f_\varepsilon \simeq \frac{0.4 \alpha}{3\pi}  \Bigl(\frac{q_\varepsilon}{e}\Bigr)^2  ln\frac{E_\gamma}{2m_\varepsilon} f_{\nu}
\label{qflux}
\end{equation}

Finally, using Eq.(\ref{qecrs}) and Eq.(\ref{qflux}) and taking  
$f_{\nu} = 5.8\times10^{12}$ cm$^{-2}$ s$^{-1}$ and $N_e = 2.8\times 10^{26}$ \cite{texono},
one can find that
the number of $\varepsilon-e$ events expected to be seen in the TEXONO Ge detector per day is 
\begin{equation}
\Delta N_{\varepsilon-e}\simeq 10^{18} \Bigl(\frac{q_\varepsilon}{e}\Bigr)^4 ln\frac{E_\gamma}{2m_\varepsilon}  \frac{1}{ day}
\end{equation}
 Comparing this to the upper bound on the number of events implied 
by the upper bound on the neutrino magnetic moment 
$\mu_\nu < 1.3\cdot 10^{-10} \mu_B$ \cite{texono} and assuming    $E_\gamma \simeq $ 1 MeV 
we obtain :
\begin{equation}
q_\varepsilon \lesssim  10^{-5}  e
\label{limit}
\end{equation}
for the mass region $m_\varepsilon <  $1 keV. 
This estimate  is weakly depended 
on the $\varepsilon$-mass, 
variations of the selected recoil energy interval and 
uncertainties of the $\varepsilon$-flux.
 It gives an order of magnitude for the upper limit on $q_\varepsilon$
and may be strengthened and extended to the mass region 
$m_\varepsilon >$ 1 keV by more accurate and  detailed Monte Carlo simulations.
The limit of Eq.(\ref{limit}) is valid for a $\varepsilon$-particle  lifetime 
$\tau_\varepsilon$[s] $>10^{-13}m_\varepsilon$[eV]. The attenuation of the  flux due to (electromagnetic) 
interactions in the 
shielding is found to be negligible, since for the limit of 
Eq.(\ref{limit}) the $\varepsilon$-particle mean free path in iron is $\gg$1 km, 
as compared with the 
total shield thickness of $ < 28$ m  used at the Kuo-Sheng Nuclear 
Power Station. 

The results obtained could be also used to derive direct experimental limit on the 
electron anti-neutrino electric charge:
\begin{equation}
q_{\nu_e} \lesssim 3.7 \times 10^{-12}~e
\end{equation}
which is within two orders of magnitude of astrophysical bounds \cite{pdg}. 

The statistical limit on the sensitivity of a reactor experiment searching for 
millicharged particles  is proportional to 
$q_\varepsilon^4$ and is set by the value of $q_\varepsilon$. Thus, to improve the limit of Eq.(\ref{limit}) large   target mass is required.
Because of the $1/E_e$ dependence of the $\varepsilon-e$ scattering 
cross section the detection of low recoil electron energies 
is also crucial for the improvement of sensitivity of further searches. 
A possible approach, which could satisfy both requirements, is to
 use a massive liquid argon detector. An example of such detector is the one 
 developing in the framework of the ArDM project for the search of 
Dark Matter \cite{ardm}. 
This experiment aims at  running 
almost 1 ton of effective mass detector and  is  capable  
to measure deposited energies down to  10 keV. 
 The significance of the 
 $\varepsilon$-particles discovery  with such  detector, assuming 
$q_\varepsilon = 3\cdot 10^{-6}e$, scales as 
\begin{equation}
S \simeq 10^{-2} \Bigl(\frac{M t}{n_0}\Bigr)^{1/2}
\end{equation} 
where  
$S=2(\sqrt{n_s + n_b}-\sqrt{n_b})$ \cite{bk} and  $n_s = 10^{-2} M t$, 
$n_b = n_0 M t$ are the number of detected signal and  background events, respectively.
Here $M$ is the mass of the detector in kg, $t$ is the running time in days and 
$n_0$ is the background counting rate in kg$^{-1}$day$^{-1}$ for the energy 
interval of interest.
Thus, assuming $S\gtrsim 3$, $M\simeq 10^3$ kg and $ t\simeq 10^2$ days,
a  background level 
of $n_0\lesssim 1$ event/kg/day has to be achieved  \cite{ardm}.
The experiment could also significantly improve
the recently obtained  bounds on neutrino magnetic moment \cite{texono}, \cite{neuch}, \cite{sk}
pushing it down to the region  $\mu_\nu \simeq 10^{-12} \mu_B$.  A bound   in this  
region is comparable with the
astrophysical limits \cite{pdg} and will be  of great interest for many extensions of the  
Standard Model \cite{moh}.
Note, that  $\varepsilon-e$ events could be distinguished from the 
$\bar{\nu}-e$ magnetic scattering due to the $(1/E_e)^2$ profile of 
their recoil energy spectrum, while $\bar{\nu}-e$ events have $1/E_e$ energy 
distribution.

Another experimental approach relies on the precise study of the positronium 
atom \cite{ops}.
If low-mass millicharged particles exist,
the orthopositronium ($o-Ps$), the triplet 
bound states of electron and positron, could decay ''invisibly''
into $\varepsilon\bar{\varepsilon}$ pairs since the $\varepsilon$-particles
would penetrate any type of calorimeter mostly without interaction 
\cite{atojan,mits}.
For spin $\frac{1}{2}$ millicharged $\varepsilon$-particle and for $m_\varepsilon \ll m_e$  
the best
 experimental bound  obtained by the ETH-INR experiment 
searching for $o-Ps \to invisible$  
is  $q_\varepsilon < 3.4 \cdot 10^{-5}$  \cite{bader}, which is comparable with the limit of Eq.(\ref{limit}).

The sensitivity of  experiments on invisible decay of 
orthopositronium is proportional to $q_\varepsilon^2$, thus making further 
searches of this decay mode more  interesting and  promising \cite{ops}. 
The  ETH-INR experiment 
achieved sensitivity to  $Br(o-Ps\rightarrow \varepsilon\bar{\varepsilon})$ as small as 
$4.2 \cdot 10^{-7}$ (90\% C.L.)\cite{bader}.
To intersect the PVLAS parameter space
a new experiment with the sensitivity 
improved by a factor of 100  is required.

 Recently, a method based on 
the measuring of the electric current 
comprised of the millicharged particles produced in a cavity has been 
proposed \cite{cavity}, see also  \cite{mitra}. 
The  millicharged particles, if they exist, contribute to the energy loss of the
 cavity and thus leave an imprint on the cavity's quality factor. 
Already conservative estimates performed in \cite{cavity} show that this 
experiment will
substantially bound the parameter space for $\varepsilon$-particles allowing firmly
 exclude ( or confirm) the  region suggested by the PVLAS signal.

{\bf Acknowledgments}

We thank ETH Z\"urich, the Swiss National Science Foundation, and the INR Moscow for
support given to this research. We would like to thank Prof. V. Matveev
and Prof. V.A. Rubakov for useful remarks. 
S.N.G. is grateful to ETH Z\"urich for kind hospitality and support.

\end{document}